\documentclass[journal]{IEEEtran}

\usepackage{cite}
\usepackage{amsmath,amssymb,amsfonts}
\usepackage{graphicx}
\usepackage{textcomp}
\usepackage{xtab}
\usepackage{lineno,hyperref}
\usepackage{amssymb,amsmath}
\usepackage{float}
\usepackage{subfig}
\usepackage{multirow}
\usepackage{array}
\usepackage{lscape}
\usepackage{subfloat}
\usepackage{rotating}
\usepackage{caption}
\usepackage{color}
\usepackage{relsize}
\usepackage{epstopdf}
\usepackage{threeparttable}
\usepackage{verbatim}
\usepackage{chngpage}
\usepackage{amsmath}
\usepackage{algorithm}
\usepackage[noend]{algpseudocode}
\usepackage{subfig}
\usepackage[table, dvipsnames]{xcolor}
\usepackage{multirow, booktabs}
\usepackage{threeparttable}
\usepackage{booktabs,makecell}
\usepackage{nomencl}
\usepackage{ifthen}
\usepackage{mathtools}
\DeclarePairedDelimiter\floor{\lfloor}{\rfloor}
\def\BibTeX{{\rm B\kern-.05em{\sc i\kern-.025em b}\kern-.08em
    T\kern-.1667em\lower.7ex\hbox{E}\kern-.125emX}}

\begin{document}

\title{Delayed Q-update: A novel credit assignment technique for deriving an optimal operation policy for the Grid-Connected Microgrid}

\author{Hyungjun~Park,
        Daiki~Min,
        Jong-hyun~Ryu,
        and~Dong~Gu~Choi
\thanks{(Corresponding author: Dong Gu Choi.)}
\thanks{H. Park and D. G. Choi were with the Department of Industrial and Management Engineering, Pohang University of Science and Technology, 77 Cheongam-Ro, Nam-Gu, Pohang, Gyeongbuk 37673, Rep. of Korea (e-mail: hjhjpark94@postech.ac.kr; dgchoi@postech.ac.kr).}
\thanks{D. Min was with the School of Business, Ewha Womans University, 52, Ewhayeodae-Gil, Seodaemun-Gu, Seoul, 03760, Rep. of Korea (e-mail: dmin@ewha.ac.kr).}
\thanks{J. Ryu was with the College of Business Management, Hongik University, 2639, Sejong-Ro, Jochiwon-Eup, Sejong, 30016, Rep. of Korea (e-mail: jhryu@hongik.ac.kr).}
}



\maketitle

\begin{abstract}
A microgrid is an innovative system that integrates distributed energy resources to supply electricity demand within electrical boundaries. This study proposes an approach for deriving a desirable microgrid operation policy that enables sophisticated controls in the microgrid system using the proposed novel credit assignment technique, delayed-Q update. The technique employs novel features such as the ability to tackle and resolve the delayed effective property of the microgrid, which prevents learning agents from deriving a well-fitted policy under sophisticated controls. The proposed technique tracks the history of the charging period and retroactively assigns an adjusted value to the ESS charging control. The operation policy derived using the proposed approach is well-fitted for the real effects of ESS operation because of the process of the technique. Therefore, it supports the search for a near-optimal operation policy under a sophisticatedly controlled microgrid environment. To validate our technique, we simulate the operation policy under a real-world grid-connected microgrid system and demonstrate the convergence to a near-optimal policy by comparing performance measures of our policy with benchmark policy and optimal policy.
\end{abstract}


\IEEEpeerreviewmaketitle

\section{Introduction}

A microgrid is a system that integrates distributed energy resources (DERs) to supply the local electricity demand. The DERs include several different types of resources such as renewable energy sources (RES), energy storage systems (ESS), and small dispatchable generators (e.g., diesel engines, gas turbines; DGs). In addition, in the grid-connected mode, the electricity is tradable with external power networks (e.g., utility grid). The benefits of a microgrid are as follows: (1) the utilization of RES that efficiently reduces carbon emissions, (2) an improvement local energy delivery with low investment in capacity both for generation and transmission, (3) an improvement to the reliability and resilience of the local energy supply, and (4) a reduction in the costs associated with long-term infrastructure investment~\cite{Milis:2018, Ustun:2011}. In response to these benefits, the installed capacity of microgrids is growing fast and expected to reach 8.8 GW by 2024~\cite{Mishra:2020}.

One general goal of the microgrid operation is minimizing the operation costs by generating an optimal operation of DERs' output while satisfying the constraints of the system. The optimal operation problem involves a sequential optimization process under uncertainty variables such as time-varying demand and price of electricity. Therefore, many previous studies have attempted to use methods for sequential decision problems, such as dynamic programming (DP) and stochastic programming (SP), to derive the optimal operation policy~\cite{Su:2013,Talari:2015,Nguyen:2015}. However, most proposed methods require too much computational burden for practical application. Therefore, some studies proposed meta-heuristic algorithms to derive the optimal operation policy within a reasonable time~\cite{Pourmousavi:2010,Karami:2014}. However, these studies cannot fully capture the stochastic characteristics of real system dynamics.

Responding to the limitations of previous studies, some studies have attempted to derive the optimal operation policy of microgrids based on reinforcement learning (RL) methods. RL based methods are a useful machine learning approach wherein the learning agent optimizes policies through sequential interactions with the environment without requiring the learning agent to know any information about the system dynamics. \cite{Li:2012} suggested an effective method for deriving a microgrid operation policy using a multi-agent RL and relaxing the curse of dimensionality issues. Using multi-step Q-learning, \cite{Kuznetsova:2013} developed an RL framework for autonomous multi-state and multi-criteria decision-making for medium-term scenarios of energy storage management. Similarly, \cite{Leo:2014} proposed a forecasting based multi-step ahead Q-learning algorithm to manage microgrid systems. \cite{Venayagamoorthy:2016} introduced action-dependent heuristic dynamic programming with decision tree rules and applied an evolutionary algorithm to enhance the convergence speed for finding a near-optimal control policy. \cite{Kim:2018} proposed an algorithm to derive a policy for smart energy buildings management using the tabular Q-learning method. \cite{Levent:2019} proposed a simple rule-based microgrid control policy by applying a decision tree algorithm from a Q-table derived using Q-learning. \cite{Shang:2019} derived a continuous policy for dispatching ESS using a novel RL algorithm that is combined with Monte Carlo tree search methods, reducing the size of the searching space needed to conduct multi-step bootstrapping.

More recently, following the development of deep RL (DRL) methods that combine RL and deep neural networks (DNNs), studies have suggested DRL-based methods for the optimal operation of microgrids. In \cite{Wei:2014}, a dual iterative Q-learning algorithm was proposed to derive optimal ESS management policy in smart residential energy systems of a periodic nature. \cite{franccois:2016} applied a deep Q-learning (DQL) algorithm with a novel Q-network structure to derive a microgrid management strategy. \cite{Chen:2018} derived prosumers’ electricity trading strategy using DQL in the local energy market. \cite{Ji:2019} formulated a Markov Decision Process (MDP) model for the system of the microgrid interacting with the electricity market and derived an operation policy for the system using a deep Q-network (DQN) algorithm to conduct the real-time energy management. \cite{Bui:2019} suggested a method that derives the operation strategy of a community ESS using double DQL in both grid-connected and islanded modes with different objectives.

In previous studies proposing Q-learning based algorithms~\cite{franccois:2016,Chen:2018,Kim:2018,Bui:2019}, the defined problems contain only one action, either charging/discharging of the ESS or buying/selling electricity from the external grid. However, these approaches are insufficient and cannot be used for real microgrid applications as they are unable to work problems with sophisticated action spaces. That is, in addition to determining the direction of each DER, the algorithm needs to be able to determine the levels of the direction. For example, the algorithm should determine the amount of charging/discharging electricity at the ESS. Although \cite{Ji:2019} considered integrating the sophisticated action space in the microgrid system, they assumed that the microgrid system had only one controllable DER, the ESS. However, to the best of our knowledge, only limited research has proposed DRL based methods for several controllable DERs with sophisticated controls in a microgrid system. 

Of course, adopting the sophisticated action space for controlling several DERs simultaneously can cause an issue, in that the size of the state-action pairs increases exponentially. Particularly, when Q-learning based methods are utilized, the formation of a large discrete action space is inevitable. By constructing the large discrete action space, numerous Q-values of state-action pairs must be explored sufficiently to approximate the true value, but a learning agent cannot achieve it in a reasonable time. This issue leads to the learning agent’s inability to approximate the future value in the learning phase. Particularly in the context of the microgrid operation problem, the agent fails to embody a delayed effective property. That is, the current ESS charging control initially incurs costs but its benefit will be effective in the future when the discharging control is conducted. Consequentially, the agent only follows the myopic goal instead of planning for the long-term objective. 

Although there are several typical approximation techniques in the Q-learning based approach to support the agent in adapting well to the complex system, these approaches have not been applied well in the microgrid operation problem. Some studies strived to remedy the agent's myopic perspective by adopting the multi-step ahead approximation techniques of the action-value with the Q-learning algorithm~\cite{Kuznetsova:2013,Leo:2014}. However, these techniques require prediction-based scenarios for future steps. Therefore, these techniques can only be used for simple environmental settings with a short-term planning horizon. When the size of the action space for sophisticated control of the system increases, there will still be limits that keep the techniques from being a complete solution to the issue of the agent's myopic perspective. To overcome the issue of the large action space, this study proposes a novel credit assignment technique, delayed Q-update, to derive a well-fitted operation policy under the sophisticated action space for controlling the microgrid system.

In this context, the purpose of this study is to create a technique that allows a Q-learning based algorithm to derive an optimal microgrid operation policy of a grid-connected microgrid system under the sophisticated action space.
To apply the proposed technique, we formulate an MDP, emulating the real microgrid control system, that conducts the sophisticated controls of several DERs. To demonstrate the optimality of an operation policy derived using our proposed algorithm, we conduct a simulation of our operation policy with real-world microgrid data and compare the result with an optimal operation policy derived using DP. The main contributions of this study can be summarized as follows:

\begin{itemize}
\item This is the first study investigating an algorithm for the optimal operation of a microgrid system based on a Q-learning method with sophisticated controls of several controllable DERs, which is much more applicable in the real-world microgrid system.
\item We devise the delayed Q-update technique and corresponding Q-learning algorithm that supports the learning agent in deriving a well-fitted operation policy by overcoming the delayed effective property in the microgrid operation problem.
\end{itemize}

The remainder of this paper is organized as follows. Section 2 describes the definition of our problem and formulates the MDP model, and Section 3 explains the delayed Q-update technique with the Q-learning algorithm we used. In Section 4, we provide experimental results to validate the advantages of our approach. Finally, we conclude in Section 5 by providing relevant implications and identifying directions for future research.

\section{Problem definition}\label{sec:Problem}

In this section, we introduce the mathematical models of the microgrid operation problem. The considered microgrid system involves several DERs, such as solar photovoltaic (PV), dispatchable generator, ESS, and demand response, and it is connected to the utility grid. In the grid-connected mode, the electricity that is tradable with the external large utility grid is available to fulfill surplus or shortage demand of electricity in the system. We assume that the price of electricity traded with the utility grid is determined a day before in the day-ahead market. We would like to derive an optimal operation policy of the microgrid, which can minimize the long-run operation costs under system constraints. The mathematical formulation in Subsection $A$ describes our microgrid operation problem, similar to our previous study~\cite{Choi:2019}. Then, we explain how to build an MDP model for our RL based algorithm.  

\begin{table}[!htb]
\centering
\footnotesize
\caption{Summary of notations}\label{tab:Notation}
\begin{tabular}{*{12}{p{1.3cm}p{6.5cm}}}
    \hline
    \multicolumn{2}{l}{\textbf{Indices}}\\
    {$t$}&{index for time}\\
    \multicolumn{2}{l}{\textbf{Parameters}} \\
    ${\Delta}t$ & unit time period (e.g., hour) \\
    {$T$}&{length of planning horizon}\\
    {$M$}&{the number of episodes}\\
    {$c^{DG}$}&{dispatchable generation unit cost (e.g., KRW/kWh)}\\
    {$c^{DR}$}&{load curtailment unit cost (e.g., KRW/kWh)}\\
    {$c^{B}$}&{discharging unit cost of the ESS (e.g., KRW/kWh)}\\
    {$C^{T}$}&{transmission capacity between the microgrid and the utility grid (e.g., kW)}\\
    {$P^{DG}_{max}$}&{capacity of the dispatchable generation (e.g., kW)}\\
    {$P^{DG}_{min}$}&{minimum generation of the dispatchable generation (e.g., kW)}\\
    {$\delta^{DG}$}&{maximum ramp up \& down capacity for dispatchable generator (e.g., kW)}\\
    {$C^{B}$}&{ESS charging \& discharging capacity (e.g., kW)}\\
    {$S^{B}$}&{ESS storage capacity (e.g., kW)}\\
    {$\phi$}&{controllable demand rate (\%)}\\
    {$\rho$}&{efficiency of the ESS (\%)}\\
    \multicolumn{2}{l}{\textbf{Variables}}\\
    {$x_t$}&{state of charge of the ESS at time $t$ (e.g., kWh)} \\
    {$p_t$}&{electricity price of the utility grid at time $t$ (e.g., KRW/kWh)} \\
    {$P^{D}_t$}&{electricity demand at time $t$ (e.g., kWh)}\\
    {$P^{R}_t$}&{renewable generation at time $t$ (e.g., kWh)}\\
    {$P^{G}_t$}&{amount of electricity traded with the utility grid at time $t$ (e.g., kWh)}\\
    {$P^{DG}_t$}&{amount of electricity generated by the dispatchable generation at time $t$ (e.g., kWh)}\\
    {$P^{B}_t$}&{amount of electricity charged or discharged in the ESS at time $t$ (e.g., kWh)}\\
    {$P^{DR}_t$}&{amount of curtailed electricity demand at time $t$ (e.g., kWh)}\\
    \hline
\end{tabular}
\end{table}


\subsection{Mathematical modeling}\label{subsec:MATH_model}

The mathematical model for our microgrid operation problem is described by Equations (\ref{eq:mathmodel1})-(\ref{eq:mathmodel7}), and all notations for the model are summarized in Table~\ref{tab:Notation}.
\begin{align}
\min & \sum_{t=1}^{T}C_t(s_t,a_t) \text{,} \label{eq:mathmodel1} \\
\textrm{s.t.} \hspace{0.1cm} &  P^D_t=P^G_t+P^{DG}_t+P^B_t+P^{DR}_t+P^R_t \hspace{0.5cm} \forall_t  \text{, } \label{eq:mathmodel2}\\
&  P^{DG}_{min} \leq P^{DG}_t \leq P^{DG}_{max} \hspace{0.5cm} \forall_t  \text{,} \label{eq:mathmodel4}\\
&  |P^{DG}_t-P^{DG}_{t-1}| \leq \delta^{DG}{\Delta}t \hspace{0.5cm} \forall_t  \text{,} \label{eq:mathmodel5}\\
-\min &  \Biggl\{\frac{(S^B-x_t)}{\rho},C^B{\Delta}t\Biggr\}\leq P^B_t \leq \min\{{\rho}x_t,C^B{\Delta}t\} \hspace{0.1cm} \forall_t  \text{,} \label{eq:mathmodel6}\\
&  0 \leq P^{DR}_t \leq {\phi}P^{D}_t \hspace{0.5cm} \forall_t  \text{,} \label{eq:mathmodel7}
\end{align}

Equation (\ref{eq:mathmodel1}) represents the objective of our microgrid operation problem, minimizing total operation costs over the planning horizon. The operation costs of each time step can be defined as below (Equation (\ref{eq:math1})): 
\begin{align}\label{eq:math1}
& C_t(s_t,a_t)=c^{DG}P^{DG}_t+c^B[P^B_t]^++p_tP^G_t+c^{DR}P^{DR}_t \text{,} 
\end{align}
where $[\cdot]^+$ denotes a non-negative value (i.e., $\max\{\cdot,0\}$). All other equations describe the system constraints. Equation (\ref{eq:mathmodel2}) represents a balance constraint between the supply and demand of electricity. As constraints on dispatchable generators, Equation (\ref{eq:mathmodel4}) represents the upper/lower bound of the generation of these resources and Equation (\ref{eq:mathmodel5}) represents that the difference with previous outputs of the resources is limited in a unit time (i.e., ramp-up/down constraint of dispatchable generators). In Equation (\ref{eq:mathmodel6}), upper bound means that the amount of electricity to be discharged from the ESS is limited by a certain capacity and also by the current state of charge (SOC) level with considering on efficiency rate. Meanwhile, the lower bound of the equation means that the amount of electricity to be charged from the ESS is limited by a certain capacity and also by the remaining capacity of the ESS storage with considering on efficiency rate. Equation (\ref{eq:mathmodel7}) implies that the amount of demand response must positive and lower than the maximum controllable demand. Although the maximum transmission capacity from/to the utility grid is constrained, we can ignore the constraint in that the capacity is large enough to transmit the required amount in the problem.

\subsection{MDP modeling}\label{subsec:MDP_model}

In our microgrid operation problem, the RL agent can consider five key features in the microgrid system. The first feature is the output of the dispatchable generators in the previous hour (i.e., $P^{DG}_{t-1}$), which is used for the ramp-up/down constraints (i.e., Equation (\ref{eq:mathmodel5})). Second and third are the output of renewable energy generators and the amount of demand currently occurring (i.e., $P^R_t, P^D_t$). Finally, the current SOC level and electricity price of the utility grid are needed for configuring the current situation of the microgrid (i.e., $x_t, p_t$). In summary, the state space of the RL agent can be represented as follows (Equation (\ref{eq:math3})).

\begin{align}\label{eq:math3}
& s_t=(P^{DG}_{t-1},P^R_t,P^D_t,x_t,p_t) \text{,} 
\end{align}

The RL agent needs to determine the outputs of controllable DERs and the amount of electricity from/to the utility grid. As a result, the action space of the RL agent has four actions, as shown by Equation (\ref{eq:math4}). The first action is the amount of electricity from/to the utility grid. If the action has a negative value, it means that the microgrid sells the electricity to the external grid. The amount of small DGs and the demand response are also considered as actions. The last one is the charging/discharging amount of ESS. If the action has a negative value, it means that the microgrid charges electricity into the ESS. Here, the action space determines the last three elements of the action space only as the first element (i.e., $P^G_t$) is automatically determined by the balancing constraint between supply and demand (i.e., $P^{G}_{t}=P^{D}_{t}-P^{R}_{t}-P^{DG}_t-P^B_t-P^{DR}_t$). By doing this, any action selected by that agent can satisfy the first constraint (Equation~\ref{eq:mathmodel2}) and also the dimension of the control variable decreases to tractable sizes.

\begin{align}\label{eq:math4}
& a_t=(P^{G}_{t},P^{DG}_t,P^B_t,P^{DR}_t) \text{,} 
\end{align}

The reward in the MDP model should reflect how much the agent's action contributed to minimizing the microgrid operation costs. This reward can be simply defined as the negative sign of microgrid operation costs per time step (i.e., one hour). However, if a reward is defined as this simple version, then non-stationary reward criteria may be applied depending on the demand phases (e.g., the phase is categorized as the peak demand and base demand). For example, during the base demand phase (i.e., the periods with relatively low demand), no matter how undesirable the conducted action is, a relatively positive reward is obtained by the agent. In contrast, in the peak demand phase (i.e., the periods with the relatively high demand), regardless of how desirable the conducted action is, a relatively negative reward is obtained by the agent. Therefore, the reward formula should be robust under non-stationary conditions and constitute consistent criteria. 
In addition, to conduct stable learning, the scale of reward must be adjusted to belong to a reasonable range. The rescaled approach of the reward for Q-learning is formulated in a related study~\cite{Ji:2019}. Similarly, for the reward's reasonable scale, we define how desirable the current action is relative to the cost of the worst operation that the agent can do in each hour. This reward formula is similar to the reward formula of the reference~\cite{Park:2020} in that it is also the change rate relative to the baseline. The worst-case operation that the agent can do in this problem is to satisfy the residual demand remaining after the production amount of the RES is fulfilled using dispatchable generators (i.e., the unit cost of dispatchable generator higher than those of other resources in the experiment). Therefore, we define the reward as the negation of change rate in the hourly operation costs relative to the costs that result from the worst action case (Equation (\ref{eq:reward})).

\begin{align}\label{eq:reward}
& r_{t}=-\frac{C_t(s_t,a_t)-c^{DG}(P^D_t-P^R_t)}{c^{DG}(P^D_t-P^R_t)} \text{, }
\end{align}

The system dynamics of the MDP model only consider the SOC level of the ESS because the other elements of state do not depend on the actions in the previous time step. Equation (\ref{eq:dynamic_soc}) represents the dynamics of SOC level dependent on the action $P^B_t$. In the equation, SOC level decreases by $P^B_t$ multiplied with the efficiency rate when the ESS is discharged, otherwise SOC level increases by $P^B_t$ divided by the efficiency rate when the ESS is charged.

\begin{align}\label{eq:dynamic_soc}
& x_{t+1}=\begin{cases} x_{t}-\frac{P^B_t}{\rho} \hspace{0.2cm}if \hspace{0.2cm}P^B_t\geq0, \\ x_{t}-\rho{P^B_t} \hspace{0.2cm}if \hspace{0.2cm} P^B_t < 0 \end{cases} \forall_{t} \text{, }
\end{align}

\section{Methodology}\label{sec:Methodology}

In this section, we introduce our proposed approach for deriving the optimal operation policy of a microgrid using Q-learning with a novel technique, delayed Q-update. We leverage the advantage of the Q-learning algorithm to keep the agent from being trapped in local minima unlike other policy-based RL algorithms~\cite{Bui:2019}. 

\subsection{Description of Q-Learning Agent Model}\label{subsec:Agent modeling}

Q-learning is a value-based RL method that approximates an action value (i.e., a Q-value) in each state-action pair. The Q-value represents an expected value of a discounted cumulative reward, initiated with the current state and selected action under a policy (Equation (\ref{eq:background1})). The optimal Q-value is induced by an optimal policy, which maximizes the Q-value of every state (Equation (\ref{eq:background2})), and this value is expressed using the Bellman equation (Equation (\ref{eq:background3})). In Q-learning, the optimal Q-value can be approximated by bootstrapping, meaning that the target value of Q-value is utilized from the previous approximated value, the so-called temporal difference learning (Equation (\ref{eq:background4})). Here $\alpha$ denotes the learning rate and $\gamma$ denotes the discount factor for discounting time value. As revealed in Equation (\ref{eq:background4}), this algorithm is a model-free method such that even if the agent does not have knowledge of the environmental factors such as transition probability, the agent can develop a policy using repeated experience by following an behavior policy (i.e., epsilon-greedy). In addition, Q-learning is an off-policy algorithm, that is, the behavior policy for selecting the agent's action is not the same as the target policy for selecting an action on the target value.

\begin{align}
& Q_{\pi}(s_t,a_t)={\mathbb{E}}_{\pi}\bigg[{\sum}_{k=0}^{\infty}{\gamma}^kr_{t+k}|s_t,a_t\bigg]  \text{, }\label{eq:background1} \\
& Q^*(s_t,a_t)=\underset{\pi}{\operatorname*{max}}\hspace{0.1cm}Q_{\pi}(s_t,a_t)  \text{, }\label{eq:background2}\\
& Q^*(s_t,a_t)=\underset{a'}{\operatorname*{max}}\hspace{0.1cm}[r_t+{\gamma}Q(s_{t+1},a')]  \text{, }\label{eq:background3}\\
& Q(s_t,a_t){\leftarrow}(1-\alpha)Q(s_t,a_t)+\alpha\bigg(r_t+{\gamma}\underset{a'}{\operatorname*{max}}Q(s_{t+1},a')\bigg)  \text{, }\label{eq:background4}
\end{align}

When adopting a Q-learning algorithm, we should define the discrete range of the agent's state and action spaces. For the discretization of the agent's action space, we define each control variable divided by 10 units. By following this discretization rule, the amount of dispatchable generators contains $(P^{DG}_{max}-P^{DG}_{min})/10+1$ discrete cases which are achieved by dividing the possible range ($P^{DG}_{min}{\sim}P^{DG}_{max}$) by 10. Likewise, the charging/discharging amount of the ESS is discretized to $(S^{B}+C^{B})/10+1$ cases by dividing the possible range ($-S^{B}{\sim}C^{B}$) by 10 units, and the amount of the demand response is discretized to $\floor{max({\phi}P^D_t)/10}+1$ cases by dividing the possible range ($0{\sim}max({\phi}P^D_t)$) by 10 units. The amount of electricity from/to the utility grid is determined when all other control variables are fixed, so it does not need any consideration. Unlike the action space, the discretization for the state space cannot be directly applied because the observed features of the system cannot be exactly divided into a discrete unit. The issue of the state space discretization can be resolved by introducing function approximation-based Q-learning, and our proposed method can be extended to the research line of function-approximated Q-value. Since tabular-based Q-learning is utilized in this study, we had to conduct a process of relaxation for the system environment, which makes the system tractable using tabular Q-learning and we describe the details in Section~\ref{subsec:setting}.

As we defined in Section~\ref{subsec:MATH_model}, our handled system has several constraints. Therefore, our agent model has a constrained action space. Although RL with a constrained action space is a tricky issue and has recently been the focus of much research, this topic is not the focus of this study. Therefore, we tackle the constraint action space of our agent by adopting the simple rule of taking an action that has the highest Q-value among the feasible action spaces (i.e., masking all infeasible actions). 

\subsection{Delayed Q-update technique for ESS Controls}\label{subsec:algorithm}

In previous research on applying Q-learning to operation cost minimization in microgrid systems, hourly operation costs were provided as part of an immediate reward. However, the effect of the ESS charging control in the current period is not reflected in the current immediate reward but in the value of the future period with discharging control. Although in the updated formula of temporal-difference (TD), the expected future reward may reflect the value of charging control, the approximation for the value of charging control is incorrect for the true value because TD has an issue with the biased value-function~\cite{Sutton:1994}. This bias issue occurs more seriously under the large action space with the delayed effective property of the microgrid. Thus, the values of the ESS charging control are depreciated, resulting in undesirable operations utilizing ESS operation inefficiently. This issue does not often occur in previous studies because most of them consider only simple action spaces of an agent. As the degree of the action space's fineness increases, the learning agent needs more iterations to approximate the true value of charging control. To achieve precise Q-value assignment with a reasonable number of iterations, the delayed Q-update technique is proposed to relax this issue by conducting a delayed credit assignment to the ESS charging control. 

The process of delayed Q-update technique requires a first-input-first-output (FIFO) queue for storing and tracking the history of the ESS charging amount and state-action pair in the periods with charging control. Thereafter, discharging control is conducted by utilizing stored electricity and the adjusted value is assigned to the corresponding previous charging actions. This adjusted value is proportional to the deviation between the electricity price in the previous charging period and the price in the current discharging period. The deviation of prices between these two periods measures how low the price in the charging period and how high the price in the discharging period is. Thus, the deviation can be an indicator evaluating the desirability of ESS charging/discharging pairs (i.e., a desirable policy entails conducting ESS charging in the period with the high price and discharging in the period with the low price). In this technique, we assume that the earlier charged electricity amount has a higher priority of discharging, which is why we employ the FIFO queue. Equation (\ref{eq:criteria}) represents the updating formula of the Q-value of charging control in the previous period $\tau$. Here, $P^B_{t,\tau}$ represents the charging amount at period $\tau$ which is utilized by the discharging control at period $t ({\geq}\tau)$ and $\beta$ denotes the adaptation rate for the delayed Q-update. The detailed flow of the delayed Q-update technique with the Q-learning is described in the algorithm~\ref{alg:Delayed update}. The time complexity of the proposed algorithm is $\mathcal{O}(T^2M)$, and is somewhat more complex than that of the original Q-learning algorithm (i.e., $\mathcal{O}(TM)$). However, as mentioned in Subsection $A$, the control variable of the ESS operation is discretized by 10 units and the charging/discharging controls are bounded by a certain amount of the capacity, so the time complexity of the proposed algorithm is reduced to $\mathcal{O}(TM)$, which is the same as the Q-learning algorithm.

\begin{align}\label{eq:criteria}
& Q(s_\tau,a_\tau){\leftarrow}Q(s_\tau,a_\tau)+\beta\gamma^{t-\tau}(p_t-p_\tau)P^B_{t,\tau} \hspace{0.3cm}\forall_\tau\text{, }
\end{align}

\begin{algorithm}
\caption{Q-learning algorithm with Delayed Q-update}
\label{alg:Delayed update}
\begin{algorithmic}[1]
\State Initialize Q-function: $Q(s,a)$
\State Initialize FIFO Queue: $D$
\For{episode $k=1,M$}
    \State receive initial observation state $s_1$
    \For{$t$=1,$T$}
        \If{with probability $\epsilon_k$}
            \State select random $a_t{\in}A_{s_t}$
        \Else
            \State $a_t{\leftarrow}\underset{a{\in}A_{s_t}}{\operatorname*{argmax}} Q(s_t,a)$
        \EndIf
        \State execute ${a}_t$ and observe $r_t$ and $s_{t+1}$
        \If{$P^B_t<0$}
            \State insert $(s_t,a_t,P^B_t)$ to $D$
        \ElsIf{$P^B_t>0$}
            \State $amt{\leftarrow}P^B_t$
            \While{$amt>0$}
                \State $(s_{\tau},a_{\tau},P^B_{\tau}){\leftarrow}D[0]$
                \State $amt{\leftarrow}amt+P^B_{\tau}$
                \If{$amt+P^B_{\tau}{\geq} 0$}
                    \State $P^B_{t,\tau}{\leftarrow}-P^B_\tau$
                    \State $D.pop()$
                \Else
                    \State $P^B_{t,\tau}{\leftarrow}amt-P^B_\tau$
                    \State $D[0]{\leftarrow}(s_\tau,a_\tau,amt)$
                \EndIf
                \State \small{$Q(s_{\tau},a_{\tau}){\leftarrow}Q(s_{\tau},a_{\tau})+{\beta}\gamma^{t-\tau}(p_t-p_\tau)P^B_{t,\tau}$}
            \EndWhile
        \EndIf
        \State $G_t{\leftarrow}r_t+\gamma\underset{a'{\in}A_{s_{t+1}}}{\max}Q(s_{t+1},a')$
        \State $Q(s_t,a_t){\leftarrow}Q(s_t,a_t)+\alpha(G_t-Q(s_t,a_t))$
    \EndFor
\EndFor
\end{algorithmic}
\end{algorithm}

\section{Experimental results}\label{sec:Experimental}

In this section, we demonstrate that the policy derived using our proposed technique with the Q-learning algorithm outperforms the benchmark operation policy derived using the original method of the Q-learning algorithm. In addition, to validate the optimality of the policy, we compare the performance of our policy with that of an optimal operation policy derived using dynamic programming. We conduct a simulation for real-world grid-connected microgrid systems using both our proposed approach and original Q-learning approach as a benchmark, and we verify that the derived operation policy is relatively superior to the benchmark policy based on the proposed performance measures.

\subsection{Experimental setting}\label{subsec:setting}

As discussed in ~\cite{Jeong:2019}, we experiment with real-world data for a campus microgrid installed in a university in South Korea. The microgrid has solar PV and ESS, and it covers some parts of electricity demand in the university library. We use the actual data on the hourly electricity demand and solar PV generation of the microgrid. In addition, we assume that the microgrid can trade electricity from/to the utility grid at the system marginal price (SMP), and we obtain the data for the hourly SMP from the website of Korea Power Exchange. The average electricity retail price in South Korea is reported at approximately 100 KRW/kWh~\cite{KEE:2018}. To consider when the microgrid utilizes the ESS actively, we make the SMP price fluctuate more by multiplying a scale-up (discount) factor to SMP price value. Lastly, we refer to several previous studies to set up the appropriate values for other parameters of our models described in Section II, as summarized in Table~\ref{tab:parameter} (the unit generation cost from the dispatchable generator in South Korea is approximately 500KRW/kWh~\cite{Song:2018} and the unit cost of the demand response is set at 200KRW/kWh~\cite{Choi:2019}).

\begin{table}[h]                           
 \centering
    \begin{tabular}{|c|c!{\vrule width 1pt}c|c|}
    \hline
     \rowcolor{lightgray} \textbf{parameter}   &   \textbf{value}    &   \textbf{parameter}   &   \textbf{value}    \\ \hline
     $c^{DG}$  &   500  &  $c^{DR}$ &   200      \\ \hline          
     $c^{B}$   &   50 &  $\delta^{DG}$  & 30   \\ \hline
     $P^{DG}_{max}$ &   60  & $P^{DG}_{min}$  &   0   \\ \hline
     $S^{B}$  &   50  &    $C^{B}$    &   50       \\ \hline
     $\rho$        &  1.0  &    $\phi$   &  0.2 \\ \hline
    \end{tabular}
    \caption{Parameters of grid-connected microgrid system}
    \label{tab:parameter}                            
\end{table}

As previously mentioned, the relaxation process is conducted to convert observed features of the system to discretized state space. In the process, we round off the hourly generation amount of PV, electricity demand, and SMP price to the one-th place. As a result, the generation amount of PV contains 4 discrete values (0,10,20,30), electricity demand contains 8 discrete values (40,50,60,70,80,90,100,110), and SMP price contains 3 discrete values (70,130,140). Furthermore, we apply the discretization rule to the microgrid system element of the state space. The the output of dispatchable generators at the previous hour, is divided similar to the discrete controls of the action space for controlling the dispatchable generators. The SOC level ($x_t$) contains the same number of discrete controls as the discretized ESS storage capacity which is divided by 10 units. Therefore, the SOC level contains $S^{B}/10+1$(=6) discrete controls, obtained by dividing the possible range ($0{\sim}S^{B}$) by 10 units. Considering the discrete values for the relaxed observations of the system, the size of state space is 3024.

By following this discretization rule for the action space of the agent, the dispatchable generators are discretized into 7 controls. Likewise, the operation of the ESS is discretized into 11 controls and the demand response is discretized into 3 controls. Thus, the size of the action space is 231, achieved by multiplying the sizes of discrete controls within every control variable.

By conducting several rounds of tuning, we fix the hyper-parameters for the learning agent including the number of episodes ($M$), learning rate ($\alpha$), adaptation rate for delayed Q-update ($\beta$), and discount factor ($\gamma$). These hyper-parameter values are 3000, 0.3, 0.00001, and 0.9, respectively. 

We derived an operation policy using Q-learning with delayed Q-update technique trained in March of 2018, a month period. After the policy training step, we validate the derived operation policy during validation period of approximately 24 hours ($2018/3/29\hspace{0.1cm}6:00\hspace{0.1cm}\sim\hspace{0.1cm}2018/3/30\hspace{0.1cm}5:00$). 

\subsection{Performance measure}\label{subsec:performance}

For the validation, we define one of the performance measures as the average cost of microgrid operation, which is the daily average cost in the valid horizon. The formula for this measure is defined in Equation (\ref{eq:avg_cost}). Here, $T_v$ denotes the last index of the valid period.

\begin{align}\label{eq:avg_cost}
& AC = \frac{1}{T_v-T}\sum_{t=T+1}^{T_v}C_t(s_t,a_t)\text{, }
\end{align}

Furthermore, to focus solely on the performance of the ESS operation, we define a benefit measure of ESS operation, which is the summation of the cost of ESS charging when ESS is charged and the opportunity cost of ESS discharging when ESS is discharged. The formula for the benefit measure of the ESS operation is defined in Equation (\ref{eq:benefit}).

\begin{align}\label{eq:benefit}
& EB = \frac{1}{T_v-T}\sum_{t=T+1}^{T_v}p_tP^B_t\text{, }
\end{align}

To identify the degree of the convergence of the updated policy, we define a convergence measure, Q-value difference, which is the root mean square error between the Q-values of the current and previous epochs. The formula for this performance measure is defined in Equation (\ref{eq:diff}). Here, $Q_k(s,a)$ denotes the Q-values of state-action pairs after $k$ iterations (epoch) are finished. The measure expresses the degree of variation between the Q-values of all state-action pairs, so it can assess the degree of convergence.

\begin{align}\label{eq:diff}
& Qdif_k = \sqrt{\frac{1}{|S||A|}\sum_{s{\in}S}\sum_{a{\in}A}\bigg(Q_k(s,a)-Q_{k-1}(s,a)\bigg)^2} \text{, }
\end{align}

We adopt a regret value for the average cost which is a relative measure for evaluating a policy that compares it with an optimal policy based on average cost measure (Equation (\ref{eq:regret1})). Applying the same concept for the ESS benefit measure, regret for the benefit measure of ESS operation is formulated in Equation (\ref{eq:regret2}). These regret values evaluate how much each measure of the given policy deviates from the optimal measure, meaning that a policy which has a large regret then the policy is undesirable.

\begin{align}
& R^{AC}_{\pi} = {\mathbb{E}}_{\pi}[AC]- {\operatorname*{min}}_{\pi}{\mathbb{E}}_{\pi}[AC] \text{, }\label{eq:regret1} \\
& R^{EB}_{\pi} = {\operatorname*{max}}_{\pi}{\mathbb{E}}_{\pi}[EB]-{\mathbb{E}}_{\pi}[EB] \text{, }\label{eq:regret2}
\end{align}

\subsection{Result}\label{subsec:result}

\begin{figure}[!htb]%
    \begin{center}
    \subfloat[]{{\includegraphics[height=50mm,width=85mm]{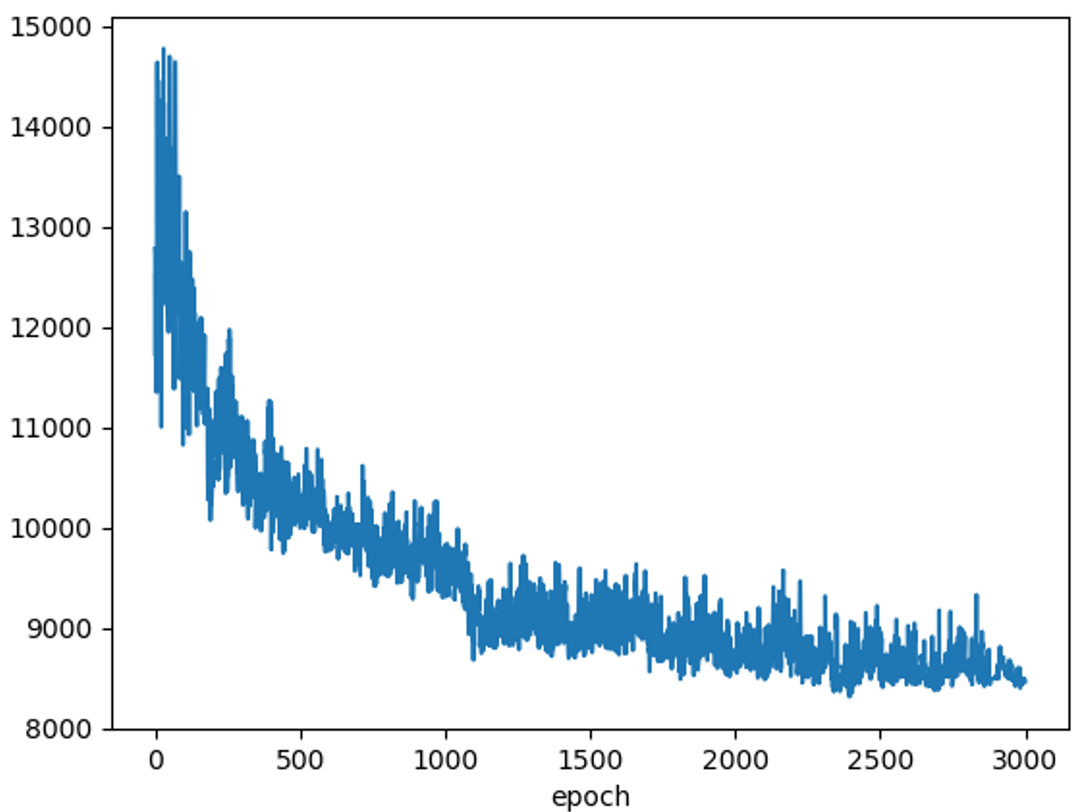}}}%
    \qquad
    \subfloat[]{{\includegraphics[height=50mm,width=85mm]{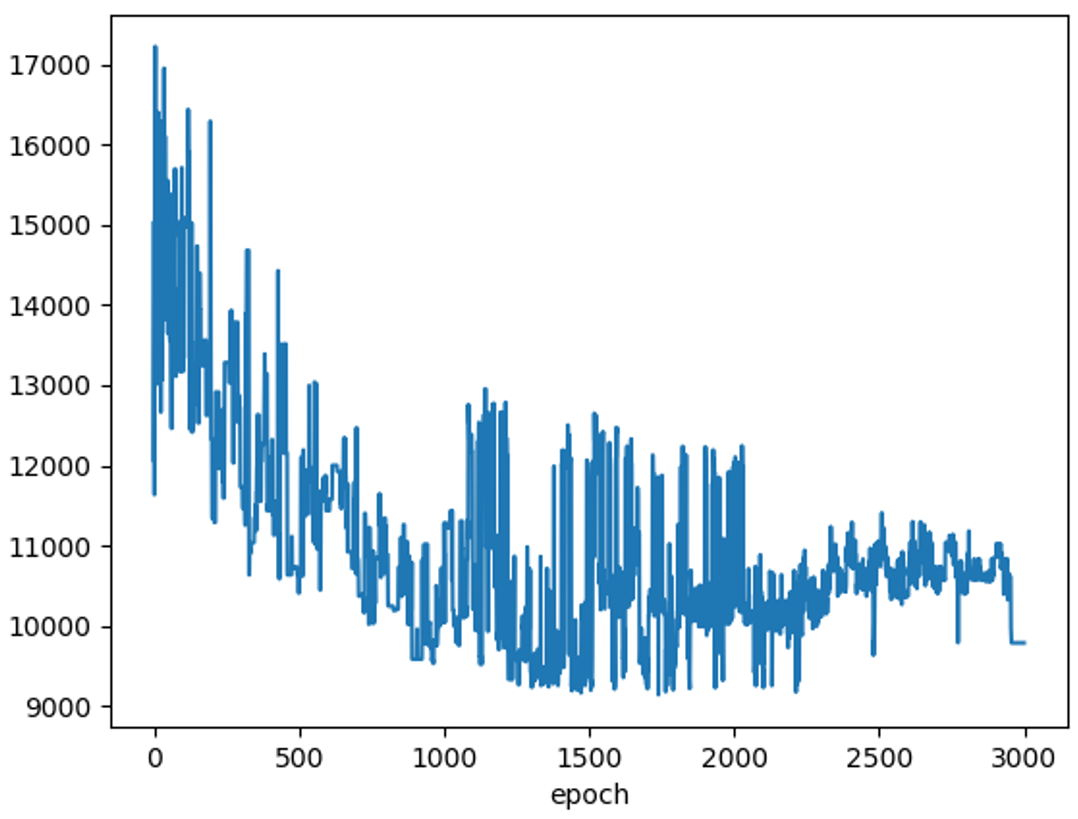}}}%
    \caption{Trend of average cost measure as the learning goes on using (a) Q-learning with delayed Q-update technique and (b) original Q-learning}\label{fig:trend}%
    \end{center}
\end{figure}

We simulate our derived operation policy using Q-learning with delayed Q-update technique in a grid-connected microgrid system. To validate the effect of the delayed Q-update technique, we conduct a comparative analysis for the operation policies derived using Q-learning with and without the delayed Q-update. Figure~\ref{fig:trend} shows the trends of the average cost plotted against operation policy updates using the proposed approach and the original Q-learning. The result of the average cost shows a declining trend in figure~\ref{fig:trend}(a), it has a stable trend as policy updating is continued before converging to a certain point. In figure~\ref{fig:trend}(b), we can identify that it has a more unstable trend than the result of our proposed method and it may not converge.

\begin{figure}[!htb]%
    \begin{center}
    \subfloat[]{{\includegraphics[height=50mm,width=85mm]{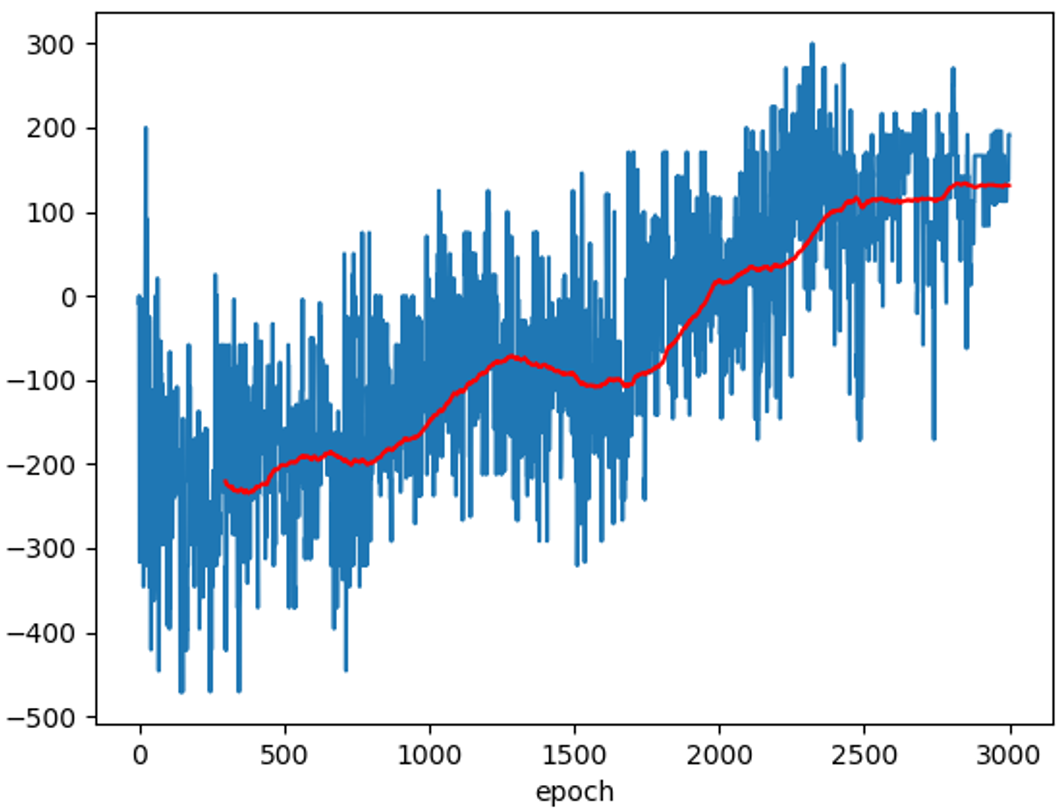}}}%
    \qquad
    \subfloat[]{{\includegraphics[height=50mm,width=85mm]{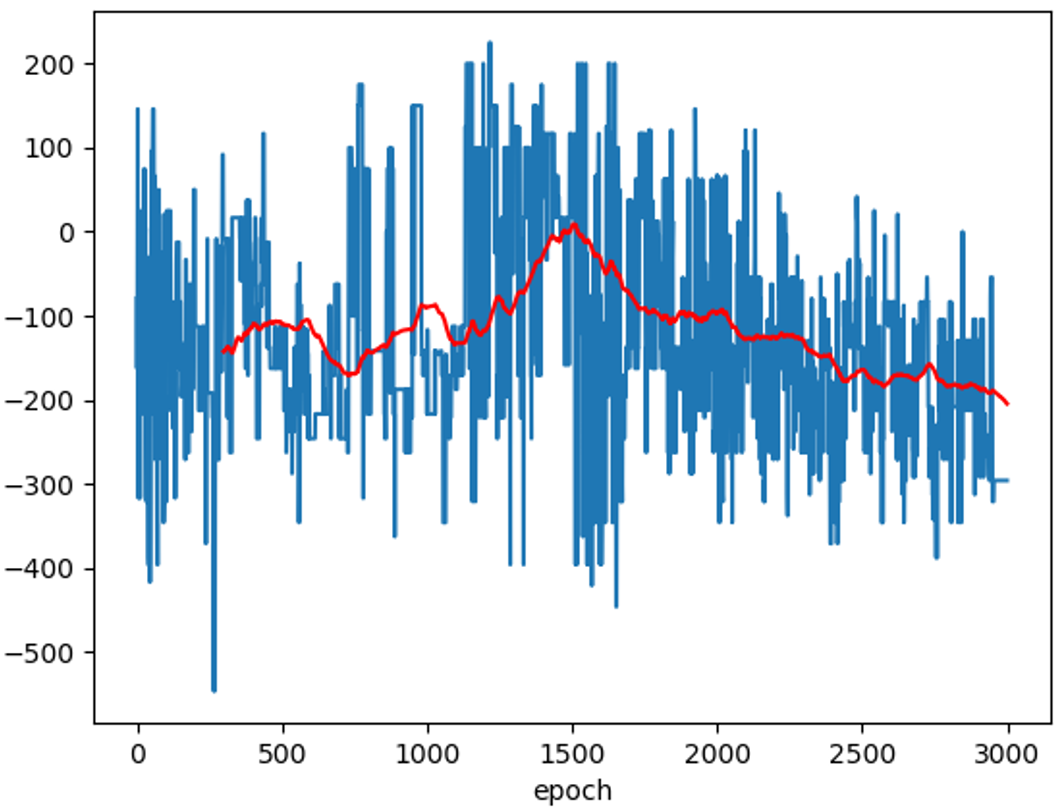}}}%
    \caption{Trend of the ESS benefit measure as the learning goes on using (a) Q-learning with delayed Q-update technique and (b) original Q-learning}\label{fig:benefit}%
    \end{center}
\end{figure}

Figure~\ref{fig:benefit} shows the trends of the ESS benefit measure versus operation policy updates using the proposed approach and the original Q-learning. To clarify the direction of each trend, we put a smoothing trend denoted by the red line in each result. As shown in figure~\ref{fig:benefit}(a), the ESS benefit measure improves as learning goes on using the proposed approach. Figure~\ref{fig:benefit}(b) shows that the ESS benefit measure does not improve and even decreases over a certain phase.

\begin{figure}[!htb]%
    \begin{center}
    \subfloat[]{{\includegraphics[height=50mm,width=85mm]{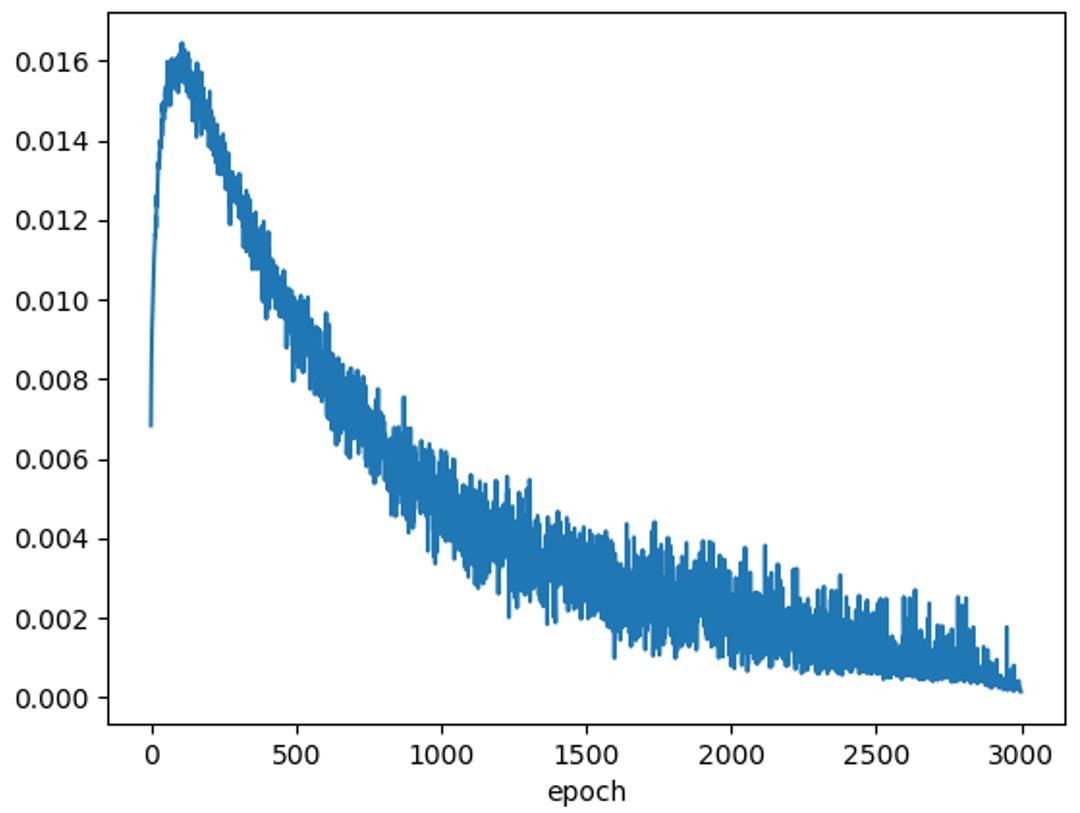}}}%
    \qquad
    \subfloat[]{{\includegraphics[height=50mm,width=85mm]{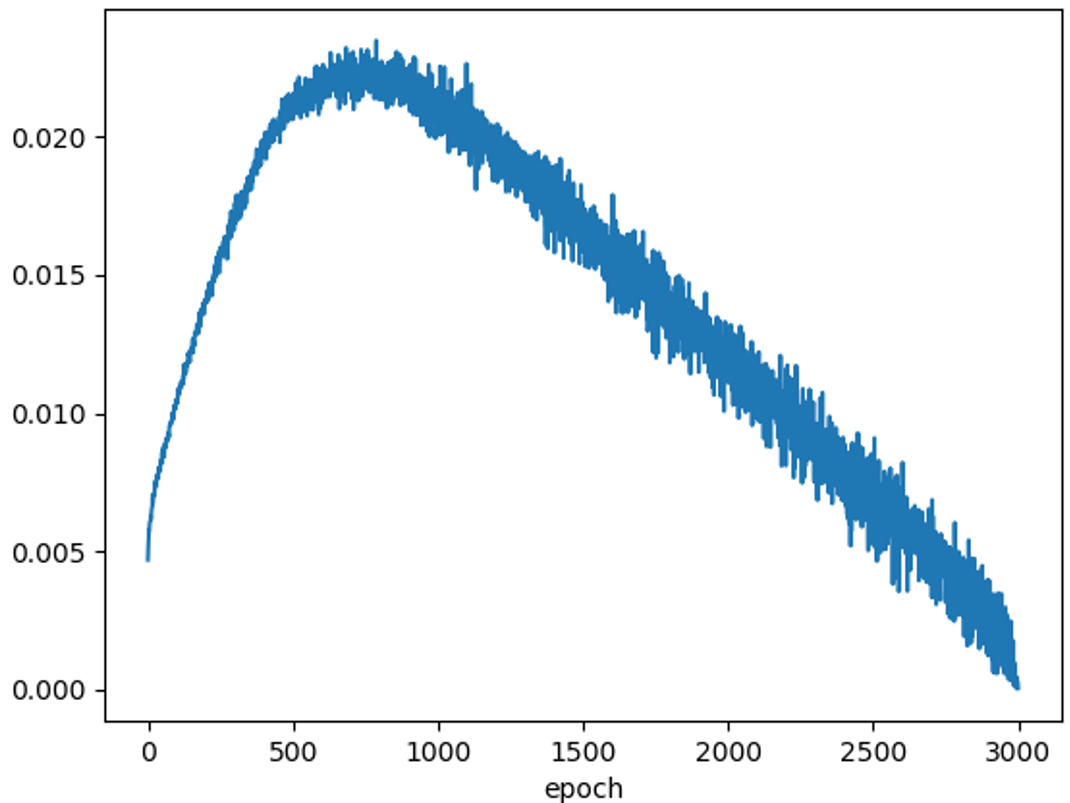}}}%
    \caption{The trend of difference measure as learning goes on using (a) Q-learning with delayed Q-update technique and (b) original Q-learning}\label{fig:diff}%
    \end{center}
\end{figure}

Figure~\ref{fig:diff} shows the trends of the Q-value difference measure plotted against operation policy updates using the proposed approach and the original Q-learning. The result of the Q-value difference shows a declining trend after a warm-up phase in figure~\ref{fig:diff} (a), it has an efficient decreasing trend as policy updating goes on before converging to a certain level. However, in figure~\ref{fig:diff} (b), we can identify that the convergence is slower and has a less efficient trend than the result of our proposed method.

\begin{figure}[!htb]%
    \begin{center}
    \subfloat[]{{\includegraphics[height=50mm,width=85mm]{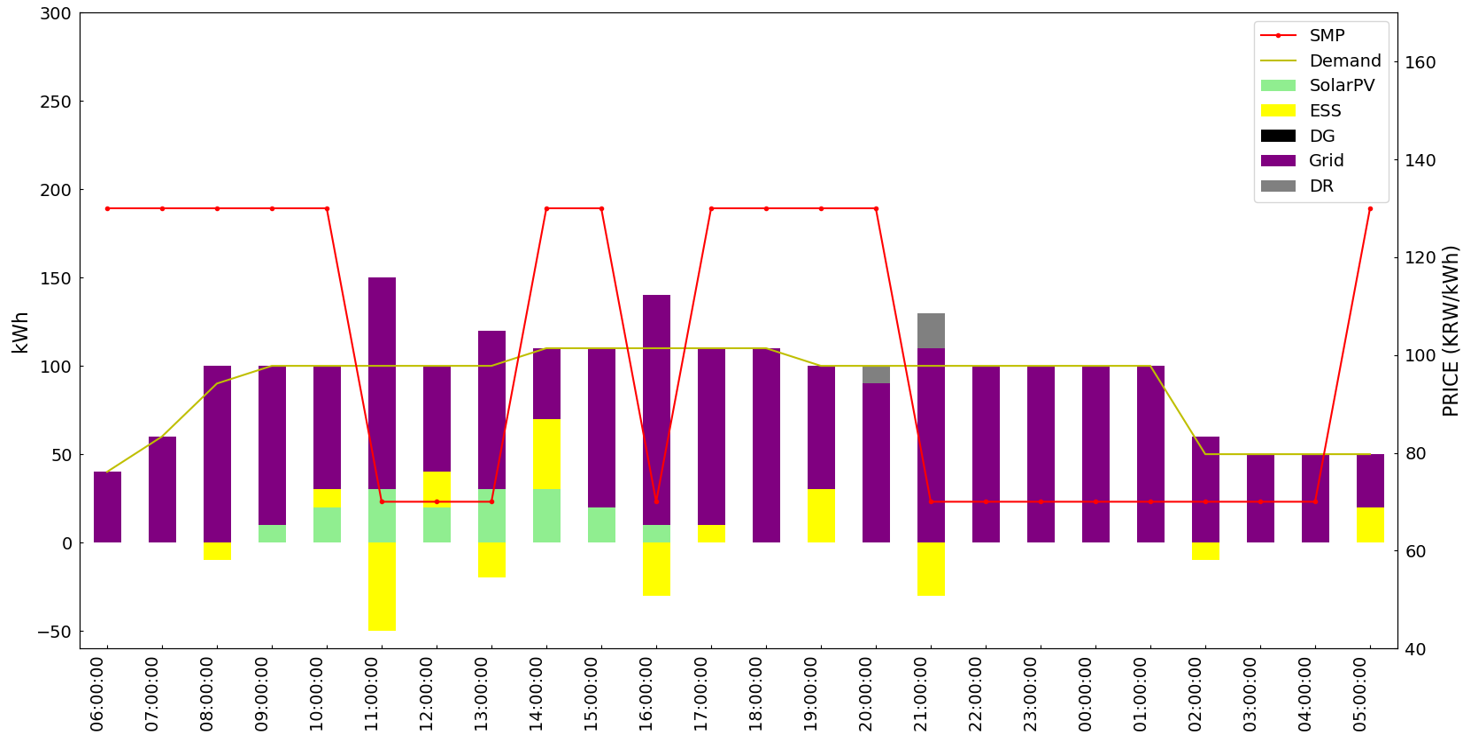}}}%
    \qquad
    \subfloat[]{{\includegraphics[height=50mm,width=85mm]{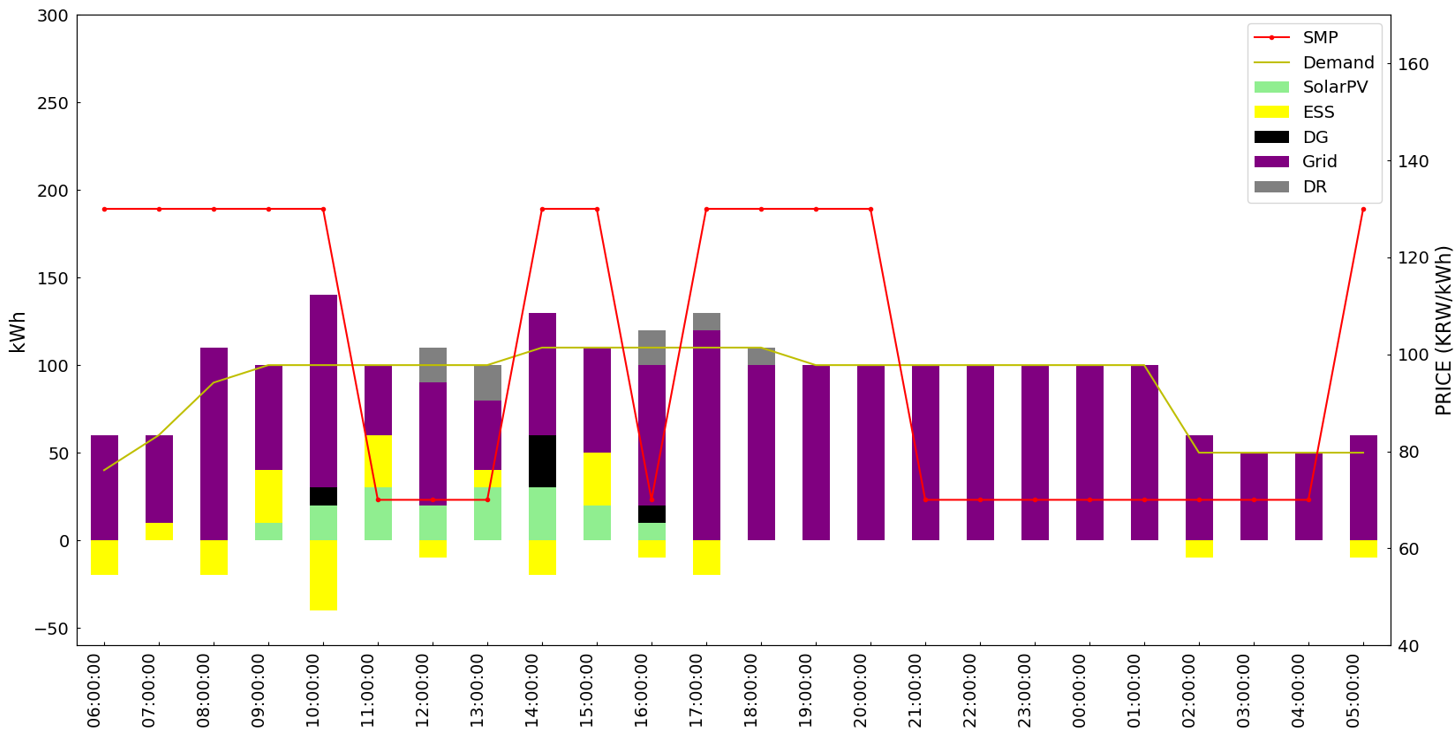}}}%
    \caption{Microgrid operation result using a policy derived by (a) Q-learning with delayed Q-update technique and (b) the original Q-learning algorithm}\label{fig:result}%
    \end{center}
\end{figure}

In the verification process of the valid data, figure~\ref{fig:result} (a) shows the operation result of the proposed approach based policy and figure~\ref{fig:result} (b) shows the operation result using a policy derived by the original Q-learning algorithm. As seen in the results, the operation from the policy derived by the original Q-learning conducts undesirable ESS control, which frequently performs ESS charging in the period with high SMP price and ESS discharging in the period with low SMP price. Therefore, our delayed Q-update technique is necessary to derive a reasonable operation policy under our problem setting.

\begin{figure}[!htb]
    \begin{center}
  \includegraphics[height=50mm,width=85mm]{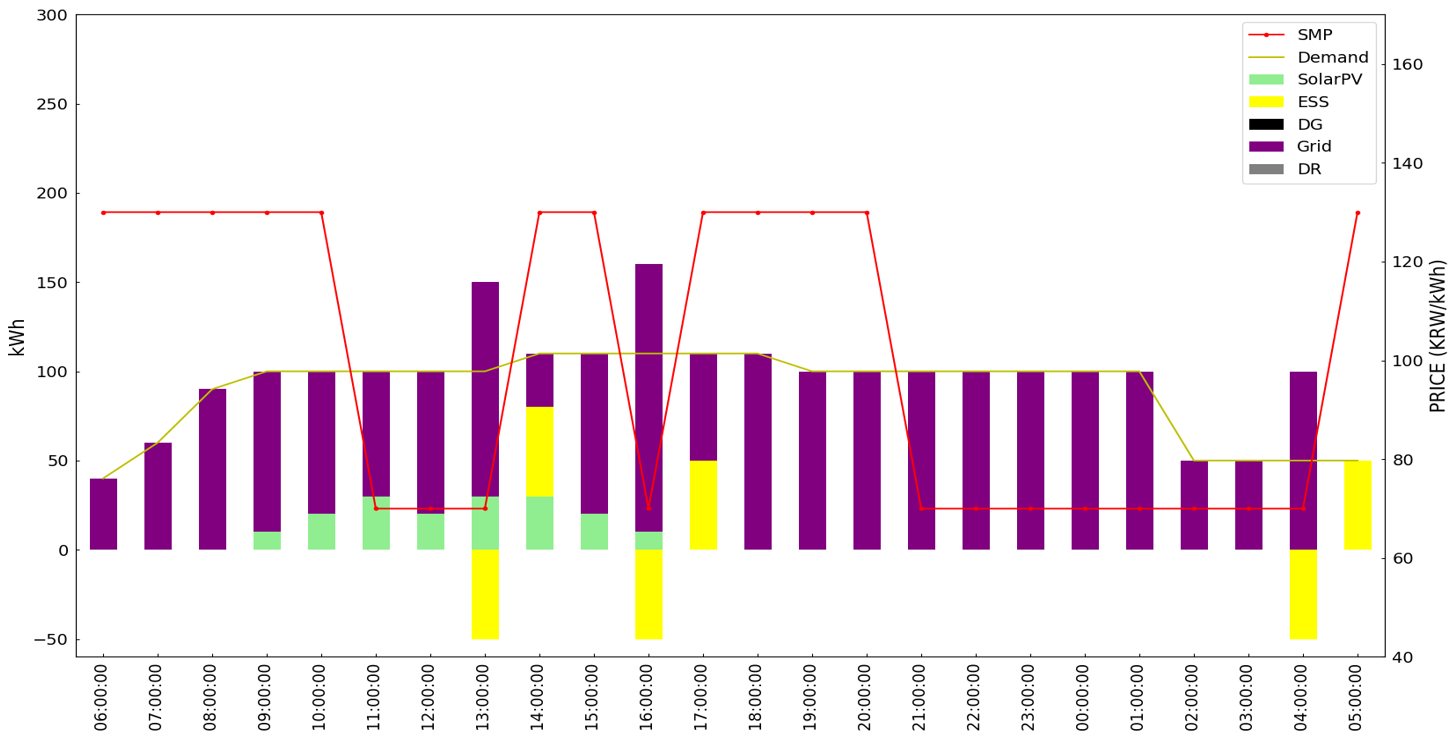}
  \caption{Microgrid operation result using optimal policy}
  \label{fig:opt_result}
    \end{center}
\end{figure}

To verify the optimality of our proposed approach, we conduct a simulation using a DP-driven operation policy (i.e., optimal policy) under the predefined system. Figure~\ref{fig:opt_result} shows the operation result, which is based on an optimal operation policy. Using an optimal operation policy, an optimal average cost is 8183.33 and an optimal ESS benefit is 375 in the valid data. Even if our operation policy somewhat differs from the optimal policy, we confirm that our policy is similar to the optimal policy in terms of charging ESS at the periods with low SMP price and discharging ESS at the periods with high SMP price. Thus, we identify that the operation policy derived using Q-learning with delayed Q-update technique converges to the near-optimal operation policy. 

To analyze the above performance measures, we summarize the final performance of the average cost of each operation policy in table~\ref{tab:vs_DQU}. As the comparative analysis shows, applying the delayed Q-update technique allows us to derive an operation policy that outperforms the policy generated by the original Q-learning algorithm. Thus, we verify that the delayed Q-update technique has a beneficial effect on the situation where sophisticated control is required under the microgrid system.

\begin{table}[!htb]                           
 \centering
\begin{tabular}{c|c|c|c|c}
  \toprule
  \rowcolor{gray!20}  & \multicolumn{4}{c}{Measure} \\ \cmidrule{2-5}
  \rowcolor{gray!20} \multirow{-2}{*}{Delayed Q-update} & average cost & ESS benefit & AC-regret& EB-regret \\
   \midrule 
   without & 9789.58 & -295.83 & 1606.25 & 670.83 \\
   with & 8462.5 & 191.66 &  297.17  & 183.34.0  \\
   \hline
   difference & -13.55\%  & 164.78\% &-81.5\%  &-72.66\%\\
 \bottomrule
\end{tabular}
\caption{Comparative analysis based on average cost, ESS benefit, and regret measures for the final derived policy under valid data between the Q-learning with the delayed Q-update technique and the original Q-learning methods}
\label{tab:vs_DQU}
\end{table}

\section{Conclusion}\label{sec:Conc}

The main contribution of our study is the creation of the delayed Q-update technique to derive operation policies more efficiently in the real-world grid-connected microgrid. Sophisticated controls are required under the real-world microgrid system, so the action space of the Q-learning agent is constructed as a large discrete action space that determines the level of every DER. There is a limitation in deriving operation policy in the large action space setting under microgrid using the typical Q-learning, especially if the ESS has a property of delayed effects, that prevents the algorithm from deriving a desirable policy efficiently. To respond with this limitation, we introduce a novel delayed credit assignment technique, delayed Q-update, which supports that the Q-learning algorithm in deriving desirable operation policies under the detailed control setting. As experimental results demonstrate, our operation policy derived using the proposed approach outperforms the benchmark policy derived using original Q-learning under adopted performance measures. In addition, these results verify that our operation policy converges to the near-optimal operation policy in the real-world grid-connected microgrid system.  

One of the limitations of this study is that the agent and environment are formulated based on the simple setting involving discrete types of electricity demand, SMP price, and generation amount of PV, adjusted by the relaxation process. However, this simple version of the environment can be extended to the real-world version which has continuous variables by adopting a recent paradigm of the RL that utilizes function approximation in Q-learning using DNN, the so-called DQN~\cite{Mnih:2013,Mnih:2015}. Along with combining DNN, to relax our derived operational policy's swift switching of the ESS controls, we plan to formulate a model that accounts for battery maintenance costs and propose a response solving approach in future research.

\ifCLASSOPTIONcaptionsoff
  \newpage
\fi



\end{document}